\documentstyle[epsfig]{mn}
\begin{document}

\title
[Multiple Lensing] 
{Strong gravitational lensing by multiple galaxies} 
\author
[Ole M\"oller and A.\,W. Blain]
{
Ole M\"oller$^{1,2}$ and A.\,W. Blain$^3$\\
$^1$Cavendish Laboratory, Madingley Road, Cambridge, CB3 0HE, UK\\
$^2$Kapteyn Institute, PO Box 800, 9700 AV Groningen, The Netherlands\\
$^3$Institute of Astronomy, Madingley Road, Cambridge, CB3 0HA, UK\\
}
\maketitle

\begin{abstract}
We discuss strong gravitational  lensing by multiple objects along any
line of  sight.  The probability  for strong gravitational  lensing by
more than  one lens is small, but  a number of strong  lens systems in
which  more than  one separate  lens contribute  significantly  to the
lensing potential will be detected in the large sample of lens systems
compiled  with  new  instruments.  Using  multi-lens  ray-tracing,  we
estimate  the likelihood for  gravitational lensing  by two  lenses at
different  redshifts  and  investigate  typical image  geometries  and
magnification  cross sections.  We  find that,  for  a cosmology  with
$\Omega_{\mathrm{M}}=0.3$  and  $\Omega_{\Lambda}=0.7$,  about one  in
twenty   lens  systems   consist  of   two  lenses   with  merging
caustics.  Multiple lens  systems  differ from  single  lenses as  the
presence of a  second lens in close proximity along  the line of sight
leads to  a strongly asymmetric potential,  which increases  the
multiple  imaging cross  section and  significantly changes  the image
configuration. The external shear induced by a second nearby galaxy,
group  or cluster can  significantly affect  image positions  even for
more  widely separated  lens  pairs.  Both of  these  effects must  be
accounted  for in lens  modelling. We  also show  how the  presence of
aligned discs  in  the  pair  of  lensing  galaxies can  lead  to  very  large
high-magnification cross sections. 
Lensing by  more than one  galaxy along the  line of
sight can lead to interesting  image configurations. Such systems will
be important in future, both for constraining lens models of individual systems and for statistical lensing. 
\end{abstract}


\section{Introduction}  
Gravitational lensing  by galaxies is usually modelled  using a single
lens.  Observational results from the  CLASS survey  (Myers et  al. 1995)  suggest a
probability $p_{\mathrm{s}}$ for gravitational  lensing of a source at
$z>1$ of about $10^{-3}$, comparing well with the theoretical prediction by Pei (1995).  
The  probability that two galaxies both lie
close enough to the line of sight to strongly lens a background galaxy
might be expected to  be roughly $p_{\mathrm{s}}^2\sim 10^{-6}$ and so
one might conclude that it is  very unlikely for a source to be lensed
strongly by two galaxies  at different redshifts. However, even though
the number of known lens systems  to date is still relatively small --
about 50 strong lens systems are known (Falco et al. 1999) -- there is
strong  evidence that  single  lens models  cannot  explain the  image
geometries and magnification ratios  in all cases. First, the observed
image  positions   and  magnifications  in  some   systems,  like  the
`Cloverleaf', require either an unreasonable large mass for the lens or a
substantial component of external shear (Kneib et al.  1998; Kneib,
Cohen \& Hjorth 2000; Soucail et al. 2000) in order
to explain  the details of the observed  image geometry. In  some cases,  the lensing
galaxy has been found  to be  part  of a  compact group  of galaxies,  and
including the potential  of the group greatly improves  the lens model
(Keeton \& Kochanek 1997; Keeton, Kochanek \& Seljak 1997; Kundi\'c  et al. 1997; M\"oller \& Natarajan
2000).  Second, it  has been  shown  recently for  two lens  systems
detected in the CLASS survey  that the strong lensing potential is due
to two  galaxies at different  redshifts, suggesting that  double lens
systems are not that uncommon (Koopmans et al. 1999; Koopmans \& Fassnacht
1999).\\ In  the next years, it  is expected that using a  new generation of
instruments, lens surveys will increase the number
of known  lens systems by at least  a factor of about  10 (Blain 1996;
2000).  The large sample of  strong lens systems should then contain a
significant number of cases in  which a background source is lensed by
two foreground galaxies.\\  Gravitational lensing by multiple galaxies
has been  investigated before by  Kochanek \& Apostolakis  (1988, KA98
hereafter) and Seitz \& Schneider (1992). In  KA98 the authors studied lensing by two  lenses at the
same  and at  different  redshifts using  a  very similar  ray-tracing
routine to that used in this  paper. However, due to  the severe limits
imposed by the  computing power available at the  time, their work was
necessarily restricted to  the study of only a  very small fraction of
the  parameter space  and  did not  include  lens evolution.  Multiple
lensing has also been   investigated  in  the   context  of
microlensing, where the lenses  are point masses (Lewis et al. 1993; Schneider, Ehlers \&
Falco  1992; Wambsganss, Witt  \& Schneider  1992). In  a cosmological
context,  multiple lensing  is naturally  incorporated  in simulations
which combine ray-tracing methods  with N-body simulations (Jain,
Seljak \& White 2000;
Wambsganss,  Cen \&  Ostriker 1998).  However, due  to  the resolution
limit  of the  N-body simulations,  such  work is  concerned with  the
lensing effect of large-scale  structure, on cluster and super-cluster
scales,  and not  with the  strong  lensing effect  due to  individual
galaxies.\\  In this paper,  we investigate  the expected  fraction of
double galaxy  lens systems  and their statistical  and characteristic
properties. In section 2 we outline the ray-tracing method and the
model of the evolving galaxy population used.  In section 3 we  show
how double  lens systems differ
qualitatively from single lens systems  in terms of both lensing cross
sections  and  image configurations.  In  section  4  we estimate  the
probability of  double lensing as  a function of source  redshift.  In
section 5  we look  at the statistical  image geometries, such  as the
ratio of quadruple:double images and discuss the expected distribution
of the masses and redshifts of the
lenses.  We   discuss  some   more  elaborate  multiple   lens  models
qualitatively in  section 6 and  look at some observational  issues in
Section 7.
\section{Method}
\subsection{Ray tracing}
The method used in this paper is based upon the ray-tracing routines
developed and described in M\"oller (1997) and M\"oller \& Blain (1998, MB98 hereafter). The statistical work in this
paper requires a large number of multi-plane ray-tracing
calculations which made it necessary to modify the routines in two ways: 
\begin{enumerate}
\item the deflection angle is calculated in two planes at different
redshifts using the full multi-lens equation (Schneider, Ehlers \&
Falco 1992). For two
lenses, the lens equation becomes:
\begin{equation}
\vec{\beta}=\vec{\theta}-\vec{\alpha_{\mathrm{A}}}\frac{D_{\mathrm{AS}}}{D_{\mathrm{OS}}}-\vec{\alpha_{\mathrm{B}}}\frac{D_{\mathrm{BS}}}{D_{\mathrm{OS}}},
\end{equation}
where $\vec{\beta}$ is the source position, $\vec{\alpha_{\mathrm{A}}}$ is the
deflection due to the near lens A, $\vec{\alpha_{\mathrm{B}}}$ is the
deflection due to the far lens B and
$D_{\mathrm{AS}}$, $D_{\mathrm{BS}}$ and $D_{\mathrm{OS}}$ are the
angular diameter distances between source plane S and lens A, lens B
and the observer respectively.
\item the code uses an adaptive grid to find all the images and determine the magnifications
and shears on both the image and source planes. In order to generate the
adaptive grid on the image plane, a coarse grid of triangles is first
lensed and a coarse magnification map is obtained, as described in
MB98. Each element on the coarse grid is then divided again into $N\times
N$ sub-elements, where $N$ is chosen to be proportional to the
magnification: $N=2+(\mu-1)/33$, for $\mu<100$ and $N=5$ otherwise. The
resulting adaptive grid is then divided into triangles and lensed. The
position of the resulting images are then found and
the magnifications calculated in the usual way. If necessary the procedure can be
repeated, but we found that a single iteration was sufficient for this
application. This improves the efficiency and achievable resolution
of the simulations by more than an order of magnitude. A different choice of $N(\mu)$ is possible, but we did not find a scaling that provides a significantly better performance than the one we used.
\end{enumerate}
In this paper we mainly consider spherical pseudo-isothermal mass distributions (PIMD) as deflectors, but the routine can deal with any parametric
spherical or elliptical lens profile.
\subsection{The lens profiles}
Since we are interested mainly in a qualitative investigation of double lensing involving two galaxy lenses we consider here only simple spherical mass distributions. Galaxy lenses are often modelled using a singular isothermal sphere (SIS). Such a model is, however, unphysical due to the infinite central surface mass density and total mass. Real galaxies are more realistically modelled using a pseudo-isothermal profile with a projected surface mass density of the form (Kneib et al. 1996, Natarajan \& Kneib 1997):
\begin{equation}
\Sigma(R)=\frac{\Sigma_0r_0}{1-r_0/r_{\mathrm{c}}}\left(\frac{1}{\sqrt{r_0^2+R^2}}-\frac{1}{\sqrt{r_{\mathrm{c}}^2+R^2}}\right),
\end{equation}
where we choose a small core radius $r_{\mathrm{0}}=0.1\,\mathrm{kpc}$ and a cut off radius of $r_{\mathrm{c}}=100\,\mathrm{kpc}$.
The total mass enclosed is then, 
\begin{equation}
M_{\mathrm{tot}}=\lim_{r\to\infty}M(r)=2\pi\Sigma_0r_0r_{\mathrm{c}},
\end{equation}
and the deflection angle at impact parameter $R$ is
\begin{equation}
\left|\vec{\alpha}_{\mathrm{A,B}}\right|=\frac{4GM_{\mathrm{tot}}}{(r_{\mathrm{c}}-r_0)Rc^2}\left[\sqrt{r_0^2+r^2}-\sqrt{r_{\mathrm{c}}^2+r^2}+(r_{\mathrm{c}}-r_0)\right].
\end{equation}
For realistic, small core radii, $r_{\mathrm{0}}\sim0.1\,\mathrm{kpc}$, the lensing cross section does not differ significantly from that of a truncated singular isothermal profile. However, since a singularity in the mass profile reduces the number of images by one, the number of images and image geometries are different in a singular and non-singular model.  
\subsection{The lens population}
\begin{figure}
\epsfig{file=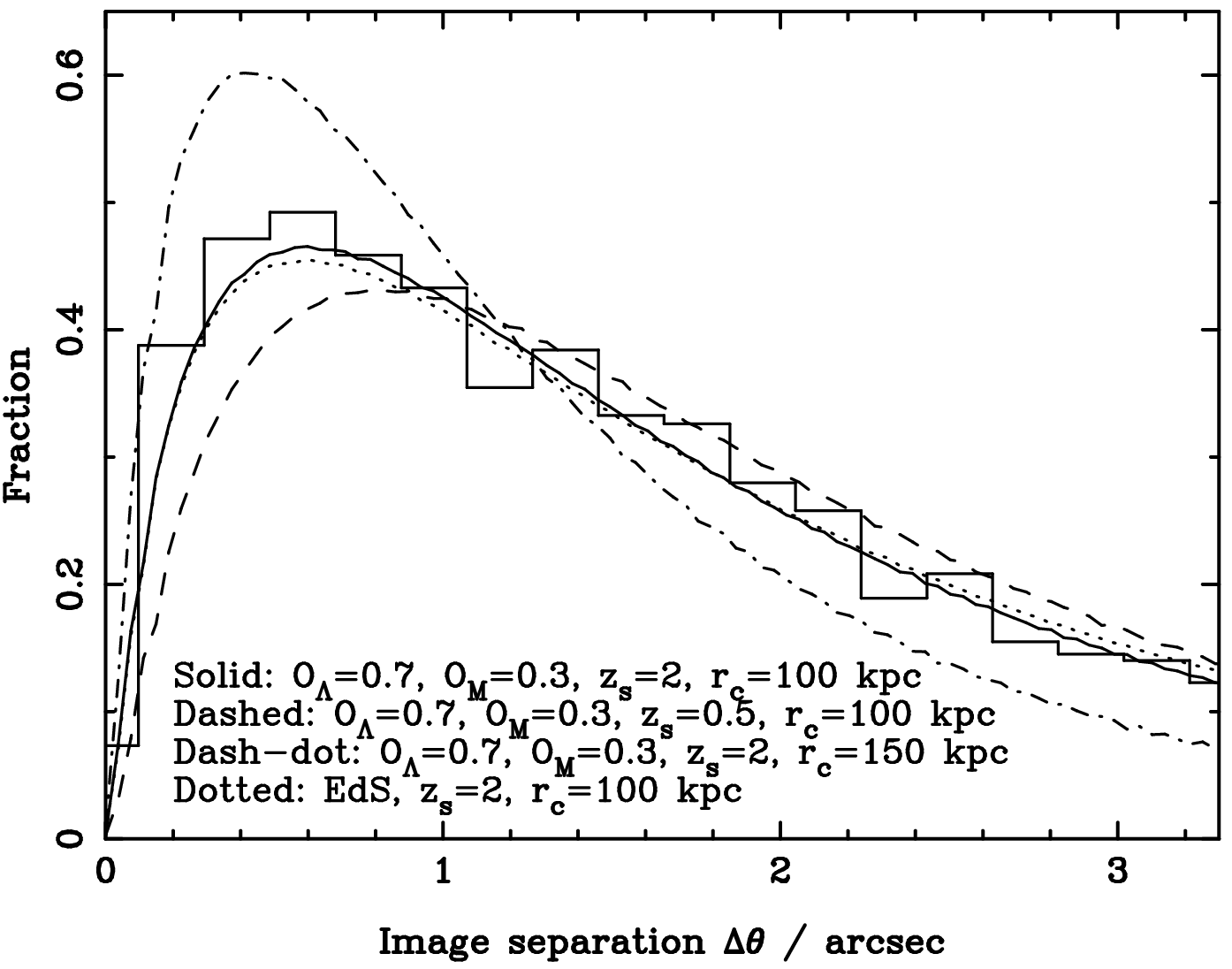,width=7.0cm,angle=-90}\\
\caption{The contribution to the total cross section for lensing of sources at
a single redshift $z_{\mathrm{s}}$ as a function
of image separation. The solid curve is
for sources at $z_{\mathrm{s}}=2$ in a  
$\Omega_{\mathrm{M}}=0.7$, $\Omega_{\Lambda}=0.3$ cosmology and a lens
populations with
$M^{*}=3.6\times10^{12}\mathrm{M}_{\odot}$, $r_0=0.1\,\mathrm{kpc}$ and
$r_{\mathrm{c}}=100\,\mathrm{kpc}$. The histogram shows
the total lensing cross section contribution toward
$z_{\mathrm{s}}=2$ in a sample of 10,000 lenses as used in the
simulation. The dashed curve shows the expected distribution
for the same lens population and cosmology but
$z_{\mathrm{s}}=0.5$. The dot-dash curve shows the expected distribution for a lens
population with $r_{\mathrm{c}}=150\,\mathrm{kpc}$ and the dotted
curve is for an Einstein-de Sitter cosmology.}
\label{contribER}
\end{figure}

In order to investigate the statistical properties of double lensing,
we created a list of 5000 lens pairs along the line of sight to a given
source at redshift
$z_{\mathrm{s}}$. The sample is determined randomly, using a simple
Monte--Carlo sampling method (Press et al. 1988) from a Press--Schechter
distribution function (Press \& Schechter 1974),
\begin{equation}
dN(M,z)=\frac{\rho}{\sqrt{\pi}}\frac{\gamma}{M^2}\left(\frac{M}{M^*}\right)^{\gamma/2}\exp\left[-\left(\frac{M}{M^*}\right)^{\gamma}\right]dM,
\end{equation}
where $\rho$ is the mean smoothed density of the universe,
$\gamma=1+n/3$ relates to the initial power spectrum index $n$, where we choose $n=1$, corresponding to a scale-invariant spectrum, and 
\begin{equation}
M^*(z)=M_0^*(1+z)^{-2/\gamma}
\end{equation}
is a characteristic bound mass at
redshift $z$.
We convert the mass from the Press--Schechter function
into a velocity dispersion for the PIMD model of the lens assuming a
cut-off radius for the lens mass distribution of
$r_{\mathrm{c}}=100\,\mathrm{kpc}$. Thus, the conversion from mass $M$
to velocity dispersion $\sigma_{\mathrm{v}}\sim\sqrt{GM/2r_{\mathrm{c}}}$.
With this conversion an $M^*$ galaxy has an approximate maximum velocity dispersion of
$245\,\mathrm{km}\,\mathrm{s}^{-1}$ if
$M^*=3.6\,\times\,10^{12}\mathrm{M}_{\odot}$ (Blain, M\"oller \&
Maller 1999). As shown in Fig.\ref{contribER} this cut-off radius is chosen so that the largest
fraction of lenses will have a separation of $\approx 1''$ as is found
in the CLASS survey (Helbig et al. 1999). For SIS lenses, and hence also, approximately, for PIMD lenses with small cores, the Einstein
radius $R_{\mathrm{E}}$ increases with the velocity dispersion $\sigma_{\mathrm{v}}$ as
$R_{\mathrm{E}}\propto \sigma^2_{\mathrm{v}}$. Note that there is a
degeneracy between the choices of $r_{\mathrm{c}}$ and $M^*$;
neither value is constrained significantly by observations. The
particular  values are chosen so that the value of $M^*$ agrees with
Blain et al. (1999).

\begin{figure}
\epsfig{file=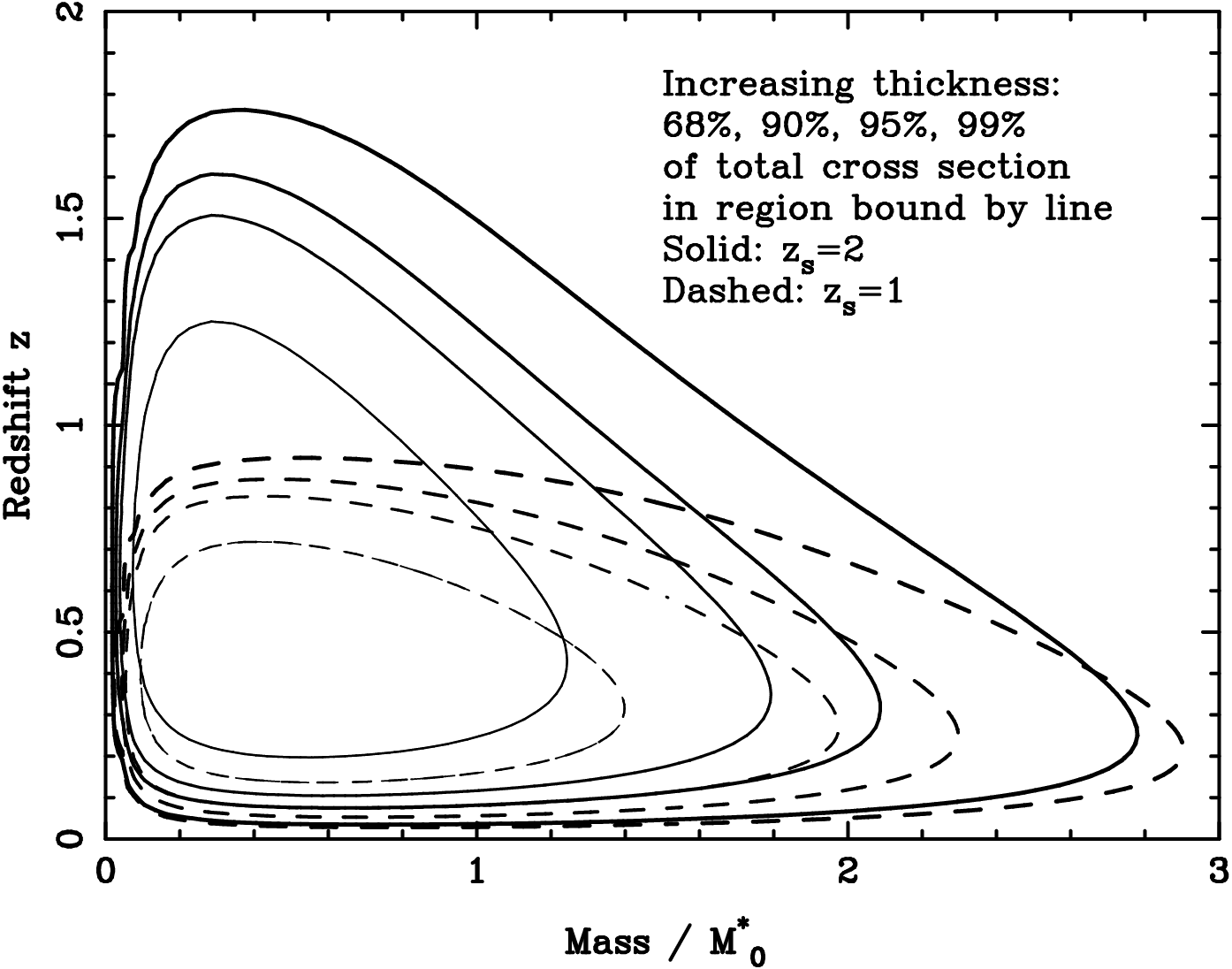,width=7.0cm,angle=-90}\\
\caption{The contribution to the total lensing cross section of
individual lenses as a function
of the mass and redshift of the lens. The lens population is given by the
Press--Schechter distribution function discussed in the text. The
source redshift $z_{\mathrm{s}}=2$ for the solid line and
$z_{\mathrm{s}}=1$ for the dashed line.}
\label{contrib}
\end{figure}

\begin{enumerate}
\item the mass of both lenses has to lie in a range $M_{\mathrm{min}}<
M <M_{\mathrm{max}}$, where the values of $M_{\mathrm{min}}$ and $M_{\mathrm{max}}$ are chosen so that less than 2 per cent of the single
lensing cross section is due to objects lying outside this range. As
shown in Fig.\,\ref{contrib}, the fraction of lenses with masses less than
$10^{13}\mathrm{M}_{\odot}$ for a value of $M^*=3.6\times
10^{12}\mathrm{M}_{\odot}$ and $z_{\mathrm{s}}=2$ is greater than 99
per cent. 
\item All lenses with an Einstein radius $R_E<0.1''$ are excluded
from the sample. As shown in Fig.\,\ref{contribER}, lenses with smaller Einstein
radii contribute less than 2 per cent to the total lensing cross
section. 
\end{enumerate}
Overall, the error on the calculated lensing cross section for double lenses that
is introduced by these selections is less than 5 per cent.
\subsection{Placement of lenses}
We choose random pairs of galaxies from the Press--Schechter
distribution function which are placed at uniformly random
positions on the lens plane inside a $4''\times4''$
field. The magnifications and image configurations for each case are
then obtained using the ray-tracing method. 
The parameters for the simulation are summarized in Table\,1.
\begin{table}
{\bf Table 1}: Parameters used in simulations.\\
\begin{tabular}{*{3}{l}}
\hline Parameter & Symbol & Value \\ \hline 
Hubble parameter & h & 0.5\\ Density parameter &
$\Omega_{\mathrm{M}}$ & 0.3 \\ Cosmological constant & $\Omega_{\Lambda}$ & 0.7\\
Critical density & $\rho$ & $2.37\Omega_{\mathrm{M}}\,h^2\times10^{11}\,\mathrm{M}_{\odot}\mathrm{Mpc}^{-3}$ \\ Initial
power-law index & $n$ & 1\\ Maximum lens mass & $M_{\mathrm{max}}$ &
$10^{13}\,\mathrm{M}_{\odot}$\\ Minimum lens mass & $M_{\mathrm{min}}$ &
$10^{10}\,\mathrm{M}_{\odot}$\\ Minimum lens redshift & $z_{\mathrm{min}}$ & 0
\\ Maximum lens redshift & $z_{\mathrm{max}}$ & 10\\ Mass parameter & $M^{*}$ &
$3.6\times10^{12}\,\mathrm{M}_{\odot}$\\ Halo cut-off radius & $r_{\mathrm{c}}$
& $100\,\mathrm{kpc}$\\
Core radius & $r_0$
& $0.1\,\mathrm{kpc}$\\ \hline
\label{parameters}
\end{tabular}
\end{table}

\section{Properties of Individual double lens systems}
\subsection{Caustics and magnification of point sources}
Two deflectors that are in close proximity to each other will modify
their respective caustic structures in a way that depends on their redshifts,
mass profiles and separation. As a first step
toward an understanding of double lensing, we used our routine to
study the qualitative dependence of the magnification maps on the parameters of such systems.\\
Panel (b) of Figs.\,\ref{mags1}, \ref{mags2}, \ref{mags3} \& \ref{mags4} show magnification maps and critical
lines for two PIMD lenses at different redshifts for four different
lens separations. The parameters for the figures are summarised in Table\,2.
\begin{figure*}
\begin{minipage}{170mm}
(a) \hskip 81mm (b)
\begin{center}
\vskip -5mm 
\hskip 9mm
\end{center}
\caption{Magnification maps and images for a double lens that consists of two PIMD
halos with total mass of $M_{\mathrm{tot}}=3\times10^12\mathrm{M}_{\odot}$. The
panel on
the left shows the results on the source plane whereas the panel on the right shows
the image plane. The
lenses have redshifts $z_{\mathrm{1}}=0.3$ and $z_{\mathrm{2}}=0.6$
and the source plane is at $z_{\mathrm{s}}=2.0$ In both panels the
angular separation of the lenses is
$\Delta\theta=1.0''$ and they are placed equidistant from the origin. The grey-scale
represents the total magnification of a source located at $x,y$ in the
source plane. The two white markers on panel (a) mark the positions
of two point sources and the single contour represents the 0.5mJy contour
of an extended source. In panel (b) the black markers show the image
positions of the two point
sources and the image of the extended source is shown as a grey
scale. The large cross marks the position of the image(s) of the more distant lens. The thick, solid lines represent the caustics, in panel (a), and the critical lines, in panel (b). The dotted lines show the individual Einstein radii of the lenses.}
\label{mags1}
\end{minipage} 
\end{figure*}
\begin{figure*}
\begin{minipage}{170mm}
(a) \hskip 81mm (b)
\begin{center}
\vskip -5mm
\hskip 9mm
\end{center}
\caption{Magnification map on the source plane (a) and images (b) for double
lenses, as in Fig.\,\ref{mags1}, but for $\Delta\theta=2.0''$.}
\label{mags2}
\end{minipage} 
\end{figure*}
\begin{figure*}
\begin{minipage}{170mm}
(a) \hskip 81mm (b)
\begin{center}
\vskip -5mm
\hskip 9mm
\end{center}
\caption{Magnification map on the source plane (a) and images (b) for double
lenses, as in Fig.\,\ref{mags1}, but for $\Delta\theta=3.0''$.}
\label{mags3}
\end{minipage} 
\end{figure*}
\begin{figure*}
\begin{minipage}{170mm}
(a) \hskip 81mm (b)
\begin{center}
\vskip -8mm
\hskip 9mm
\end{center}
\caption{Magnification map on the source plane (a) and images (b) for double
lenses, as in Fig.\,\ref{mags1}, but for $\Delta\theta=4.0''$.}
\label{mags4}
\end{minipage} 
\end{figure*}
\begin{table*}
{\bf Table 2}: Parameters for lenses in Fig.\,\ref{mags1}-Fig.\,\ref{mags4}.\\
\begin{tabular}{*{6}{l}}
\hline Parameter & Symbol & Fig.\,\ref{mags1} & Fig.\,\ref{mags2} & Fig.\,\ref{mags3} & Fig.\ref{mags4} \\ \hline 
Redshift lens A & $z_{\mathrm{A}}$ & 0.3 & 0.3 & 0.3 & 0.3\\ 
Redshift lens B & $z_{\mathrm{B}}$ & 0.6 & 0.6 & 0.6 & 0.6\\
Source redshift & $z_{\mathrm{s}}$ & 2 & 2 & 2 & 2\\
Lens separation & $\Delta\theta$ & 1'' & 2'' & 3'' & 4''\\
Position lens A & $\vec{\theta_{\mathrm{A}}}$ & (0.35'',0.35'') & (0.71'',0.71'') & (1.06'',1.06'') & (1.41'',1.41'')\\
Position lens B  & $\vec{\theta_{\mathrm{B}}}$ & (-0.35'',-0.35'') & (-0.71'',-0.71'') & (-1.06'',-1.06'') & (-1.41'',-1.41'')\\
Core radius of lenses & $r_{0}$& 0.1\,kpc & 0.1\,kpc & 0.1\,kpc & 0.1\,kpc\\
Cut off radius of lenses & $r_{\mathrm{C}}$ & 100\,kpc & 100\,kpc & 100\,kpc & 100\,kpc\\
Total lens mass & $M_{\mathrm{tot}}$ & $30\,\mathrm{M}_{\mathrm{\odot}}$ &  $30\,\mathrm{M}_{\mathrm{\odot}}$ & $30\,\mathrm{M}_{\mathrm{\odot}}$ & $30\,\mathrm{M}_{\mathrm{\odot}}$\\
\label{table_fig}
\end{tabular}
\end{table*}
These maps demonstrate four generic regimes
that occur for double lenses. It is
immediately apparent that the two lenses have a strong effect on each
other -- the magnification maps differ significantly from those of
isolated spherical lenses. In particular, there are large high-magnification
regions along the caustics.  In Fig.\,\ref{mags1} the two lenses are very close to
each other and produce a joint caustic that is extended in a direction
perpendicular to their separation. In Fig.\,\ref{mags2} the lenses are further
apart, but their individual Einstein radii still overlap. The caustic encloses a large area of high
magnification in the source plane. In Fig.\,\ref{mags3} the lenses are separated by more than
the sum of their individual Einstein radii. In this geometry the caustic is extended primarily along the
direction of the lens
separation. In Fig.\,\ref{mags4} the separation is sufficiently large that the
Einstein radii do not overlap. However, there is still a `trail' of
high magnification between the two lenses in the source plane and there is sufficient
shear to distort their individual caustics into astroids.\\ In
Fig.\,\ref{cross} we show the cross section for magnification $\mu$ of point
sources above a threshold value $A$ for the four configurations of
Figs.\,\ref{mags1}-\ref{mags4}.
\begin{figure}
\epsfig{file=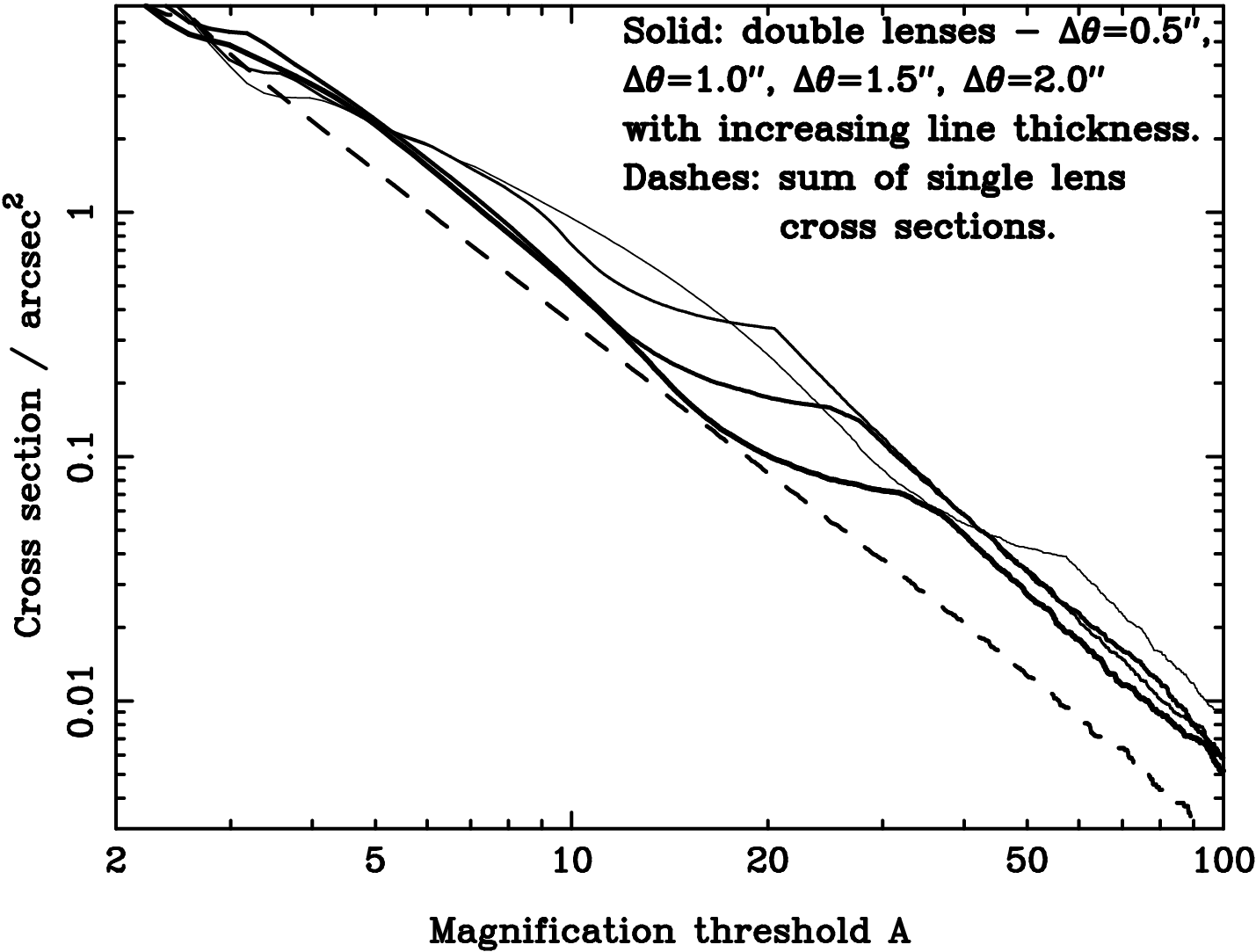,width=6.3cm,angle=-90}\\
\caption{The lensing cross section for magnification of point sources
above a threshold $\mu$. The curves correspond to the four
configurations shown in Figs.\,\ref{mags1}-\ref{mags4}.}
\label{cross}
\end{figure}
Between magnifications of $\mu\approx20$ and $\mu\approx50$ there is
an increase of up to a factor of ten as compared to the sum of the
cross sections of individual lenses. The range of magnification for which there is
a significant increase in the cross section depends on the degree of
overlap between the Einstein radii of the lenses.\\ The results 
show qualitatively that double lens systems are much more likely to
produce high magnifications of $\mu>20$ for point sources as compared
with isolated lenses. This
discrepancy between the magnification distribution for double lens systems
and the sum of the curves for the individual lenses demonstrates that
magnification bias for double lens systems is expected to be large;
hence the rough estimate of the double lensing probability as $\sim10^{-6}$ given in the
introduction is too small.

\subsection{Image geometries}

The images of point sources and one small, extended source are shown
in panels (b) of Figs.\,\ref{mags1}-\ref{mags4} for the different source positions marked in
the panel (a) to the left. In these panels, \emph{all} images are shown, irrespective of their magnification. Those images which lie in the central region of the lenses will be strongly de-magnified and will, in most cases, not be observable. Discounting those images the panels show that, due to the extended shape of the caustics, four--image geometries are common. 
Point sources that lie in the high
magnification region between the two lenses produce characteristic
aligned triple images. Extended sources in the same region produce straight
arcs with counterimages that lie on the opposite side of one of the lens centres.\\ In
small--separation 
double lens systems with different redshifts the more
distant lens can lie within the Einstein radius of the nearer lens, and so
the more distant lens can itself be multiply imaged, leading to a total of six observed lensed images. The position of the images of the more distant lens is marked by a cross in panels (b) of Figs.\,\ref{mags1}-\ref{mags4}. Also, for such small--separation systems, highly magnified sources are likely to have high image multiplicities, with three or five magnified images. For systems of intermediate separations, as shown in Fig.\,\ref{mags2}, three magnified images may appear in a peculiar triangular configuration, with large and nearly equal separations between the three images. Systems with intermediate separations are also the most likely to produce triple aligned images or straight arcs, as in Fig.\,\ref{mags3}. Large separation double lens systems, as in Fig.\,\ref{mags4}, produce image configurations very similar to those of elliptical lens galaxies.

\section{Statistics of double lensed systems}

\subsection{Definition of double lenses} 

In order to estimate the number of lens systems in which a background
source is lensed by two foreground objects, it is first necessary to
adopt a clear definition of what is meant by a ``double lens''. The
number of systems in which a second galaxy introduces only external
shear is certainly larger than the number of systems for which the background object
would be multiply imaged by both lenses individually. Here we will
define two regimes. A ``weakly coupled double lens'' is a lens system for which
the caustics for each of the two lenses do not
merge, but for which there is still a significant effect on the
individual caustic structures. A ``strongly coupled double lens'' is a lens
system in which the caustic structures merge. In other words, strongly coupled systems have a single connected multiple imaging region in the source plane, whereas weakly coupled systems do not. This distinction has
the advantage that it is easy to classify objects in each regime from the topology
of the inner caustic lines. A line connecting the two lenses will cross the high magnification
caustic line less than twice only in the strongly coupled case. For example,
the lens in Fig.\,\ref{mags4} is a weakly coupled double lens, whereas that
in Fig.\,\ref{mags2} is a strongly coupled double lens.

\subsection{Double lensing probability}

The relative probability of double lensing is given by the cross
section ratio of double lensing to lensing by the individual galaxies. The cross section due to
double lensing is given by
\begin{equation}
\int_{\vec{\mathrm{p}}}\int_0^{z_s}
\sigma(\vec{p},z_s)n(\vec{p})\frac{dV_{\mathrm{co}}}{dz}\,dz d\vec{p},
\end{equation}
where $\sigma(\vec{p})$ is the cross section for multiple images by a double
system with parameters $\vec{p}$, $n(\vec{p})$ is the comoving number
density of
such systems and $V_{\mathrm{co}}$ is the comoving volume at redshift $z$. In our model the parameters of the system are the two
redshifts, $z_1$ and $z_2$, the masses, $m_1$ and $m_2$, and the
separation of the two lenses $\Delta\theta$. The comoving number
density of objects
\begin{equation}
n(\vec{p})=n(m_1,z_1)f(m_2,z_2,\Delta\theta),
\end{equation}
where $f(m_2,z_2,\Delta\theta)$ is the probability of finding
another lens with mass $m_2$ and redshift $z_2$ at a distance
$\Delta\theta$. We assume no spatial correlation and so
$f(m_2,z_2,\Delta\theta)=f(m_2,z_2)\times f(\Delta\theta)$.
To proceed further it is necessary to compute the lensing cross
section $\sigma(\vec{p},z_s)$. Even though this is possible
analytically in the case of simple lenses (KA98), the necessary formalism is
cumbersome and cannot be extended to more elaborate lens
models. Since the ray-tracing code described above is both extremely
fast and accurate, we use it to obtain the form of $\sigma(\vec{p},z_s)$
numerically.
Instead of sampling the function $\sigma(\vec{p},z_{\mathrm{s}})$ at regular intervals,
to obtain an approximate functional form, we solve the
complete integral in equation\,5 numerically in a Monte--Carlo
fashion. This is done by sampling lenses randomly from a
Press--Schechter distribution as described in Section 2 and obtain the
value of $\sigma(\vec{p},z_s)$ for each lens system using ray
tracing.

\subsection{Numerical results}
\begin{figure}
\epsfig{file=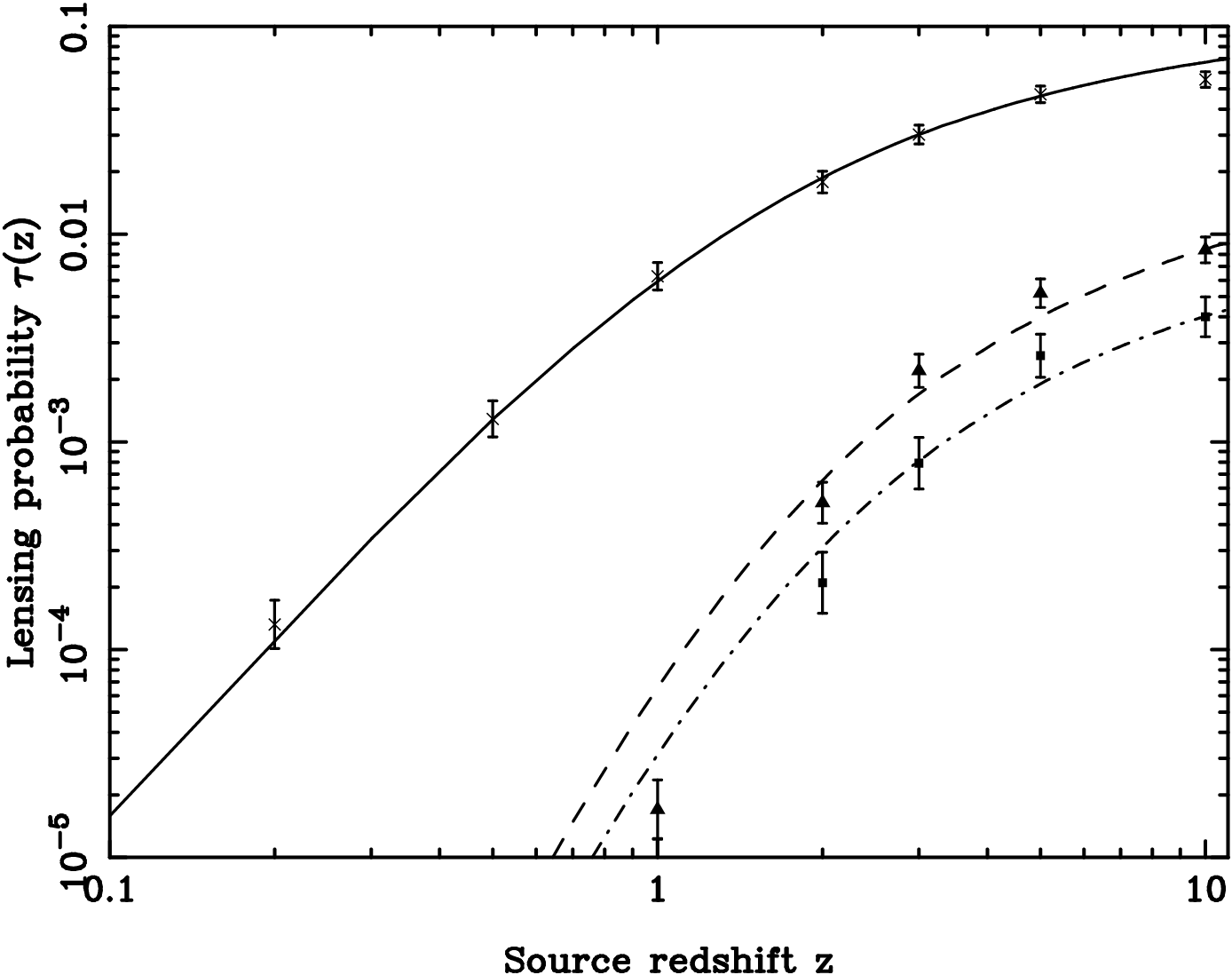,width=7.0cm,angle=-90}\\
\caption{The probability of lensing for single and double lenses. The crosses show the results from our simulations as described in the text and the
solid curve shows the analytical prediction for single lenses using
SISs with a Press--Schechter distribution
function. Squares show the probability for lensing by double lenses which have
merging caustics (``strongly coupled double lenses''). Triangles show the
probability for double lensing, including systems in which
a second lens contributes significantly to the lensing potential (``weakly coupled
double lenses''). The dashed and dot--dashed lines show the square
of the single lensing probability normalised to the numerical results
at $z_{\mathrm{s}}=10$ for weak and strongly coupled double lenses respectively.}
\label{single}
\end{figure}

Using the lens population and ray tracing lensing routines described
in Section 2, we obtain the expected number of double lens systems
numerically for ten source redshifts in the range $0.5<z_s<10$. We calculate
the total cross section for strong lensing by two lenses by summing
the cross sections for multiple imaging calculated for each lens pair
in our sample of 10,000 lenses for each source redshift, and then multiply this by the probability that two lenses
with the given properties are found inside the $4\arcsec\times4\arcsec$
field. To obtain a conservative double lensing probability, we assume that
there is no spatial correlation.\\ To test our procedure we also calculated the total multiple
lensing cross sections due to the individual lenses. Fig.\,\ref{single}. shows
the theoretical single lensing probability and the results from our
simulations for the
different source redshifts. 
As can be seen, the results agree well with
the theoretical predictions. The errors are statistical and due to the
finite  number of lenses used -- the numerical error introduced by the
ray-tracing routine is several orders of magnitudes smaller.\\ Also shown in
Fig.\,\ref{single} is the probability of double lensing as a function of source
redshift, for both weak and
strongly coupled double lens systems. The dashed and dotted curves shown are
obtained by squaring the probability for single lensing and
normalising to the double lensing probability at
$z_{\mathrm{s}}=10$.
Not surprisingly, the double lensing probability is consistently lower
than that for single lenses. However, at redshifts larger than about 2
it is above the simple estimate from the introduction. This shows
that the change in caustic structure due to the double lens potential
cannot be neglected.  Most noteworthy is
that for source redshifts of
about 5, about one in twenty lenses is expected to be a strongly coupled double
lens. Also, the number of double lens systems at redshifts
$z_{\mathrm{s}}\sim1-2$ is in good qualitative agreement with the
number of double lenses discovered so far in a radio--selected sample
(two in a sample of 50).\\
We have not assumed any correlation in our lens
sample. As the galaxy two-point correlation function is positive on
these scales (Peebles 1993), our result is therefore likely
to be an underestimate of the number of double lens systems in the
case that the two lenses are at the same redshift.
For a correlation of the form
\begin{equation}
\xi(r)=\left(\frac{r}{r_0}\right)^{-\alpha},
\end{equation}
with $\alpha\approx1.8$ and $r_0=5h^{-1}\mathrm{Mpc}$ at $z=0$ we 
estimate that the two--point correlation increases the relative number
of systems with separations below 1'' by a factor of about 4. This
means, that there will be an increase in the double lensing
probability for correlated lenses at the same redshift relative to that for uncorrelated
lenses of roughly the same factor.
As shown in Fig.\,\ref{cross} double lenses are much more likely to produce
high magnifications than individual lenses. Since flux limited lens
surveys are biased towards the discovery of highly magnified images,
the ratio of double lens systems to single lens systems in these
surveys is likely to be higher than our prediction by a factor of a few.
\section{Statistical properties}
\subsection{Lens properties}

\begin{figure*}
\begin{minipage}{170mm}
(a) \hskip 81mm (b)
\begin{center}
\vskip -5mm \epsfig{file=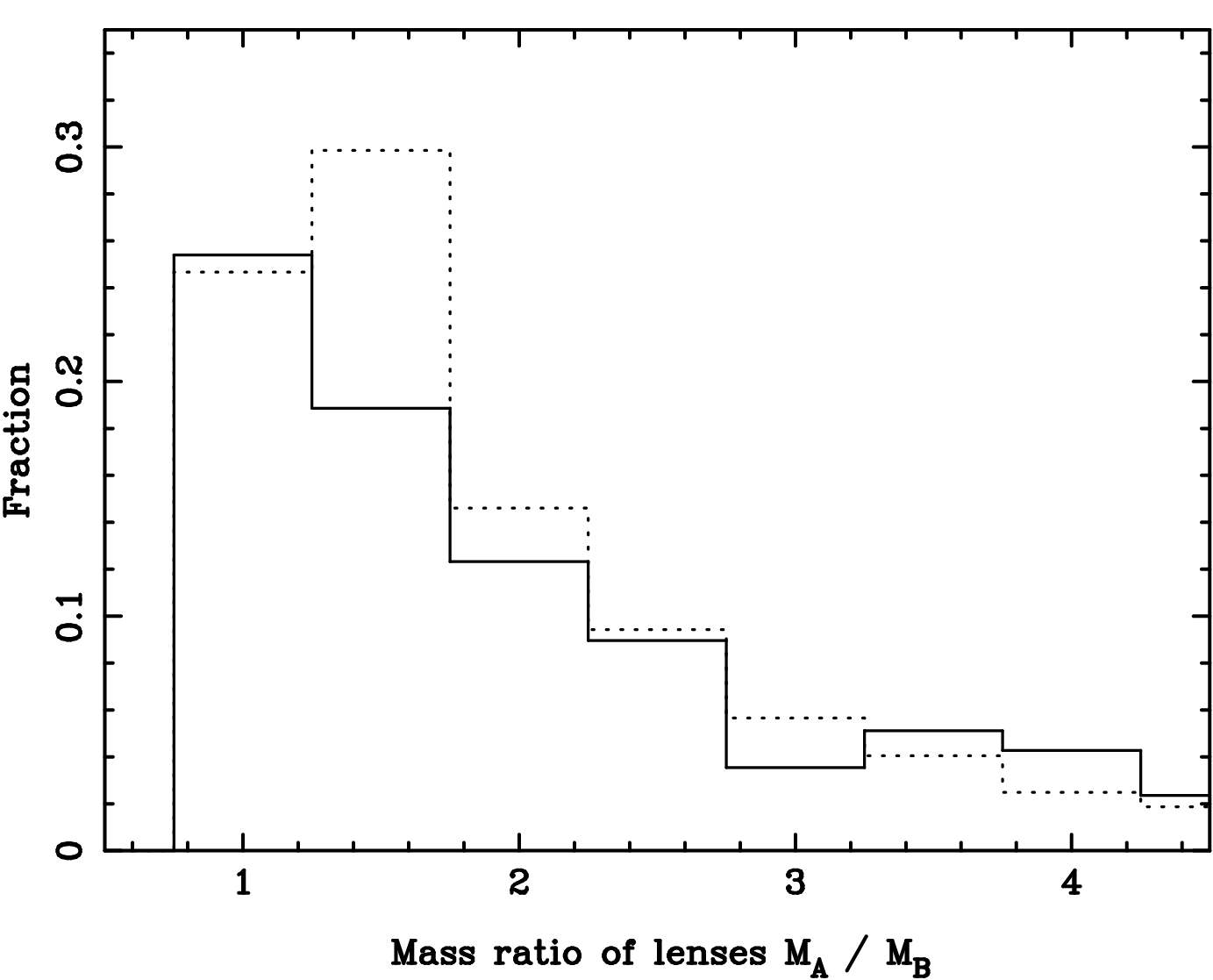,width=6.0cm,angle=-90} \hskip 9mm
\epsfig{file=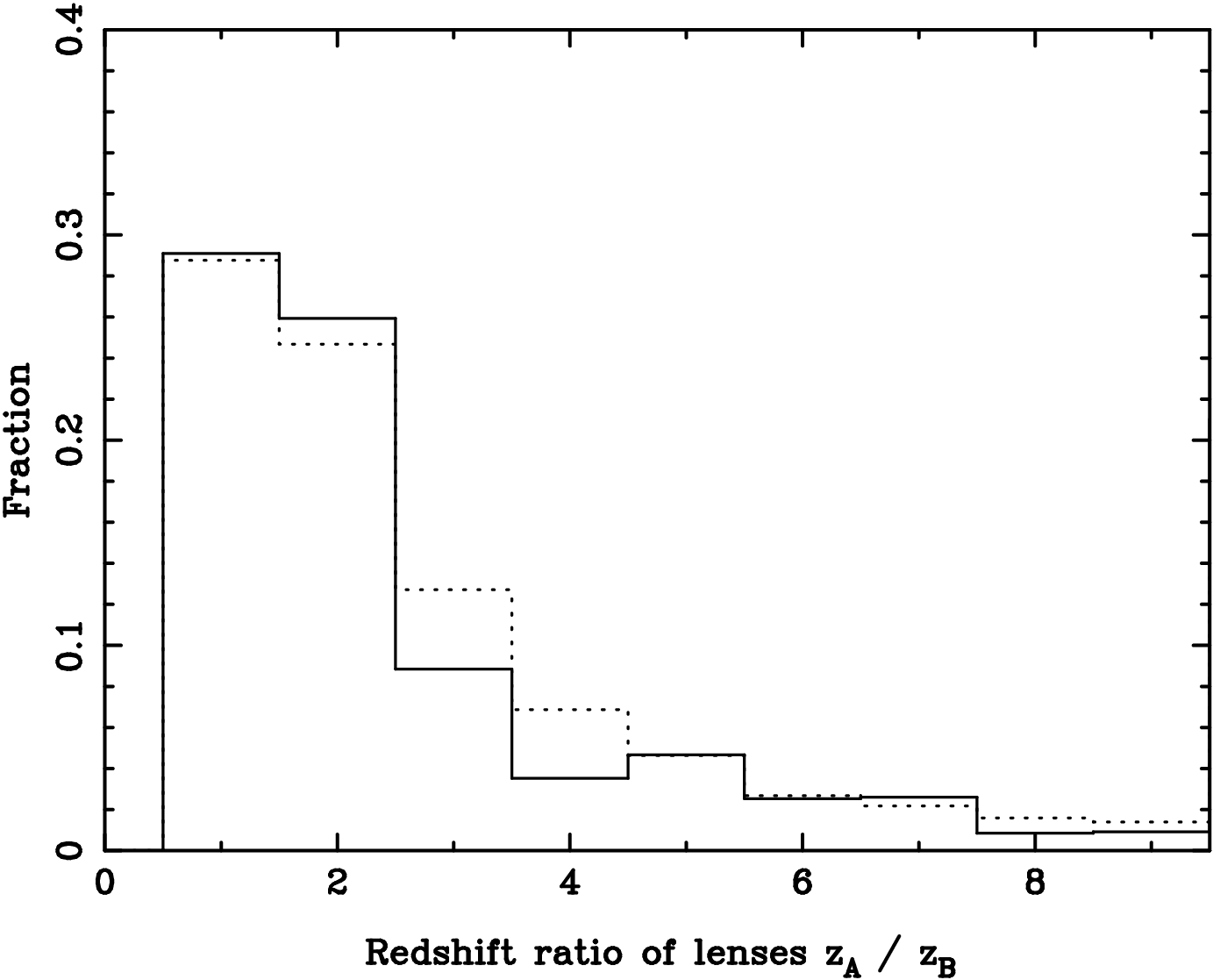,width=6.0cm,angle=-90}\\
\end{center}
\caption{Properties of double lenses from a sample of 5000 pairs,
generated as described in the text. The left panel shows the total lensing
cross section by double lenses as a function of the mass ratio. The right panel shows the
total lensing cross section as a function of the redshift ratio. The solid line
represents lens systems in which the caustics of the two lenses merge
(``strongly coupled double lenses''). The dotted line represents lens systems in which the caustics do not merge but in
which the
second lens produces significant external shear (``weakly coupled double
lenses''). The source redshift is $z_{\mathrm{s}}=2$.}
\label{masses}
\end{minipage} 
\end{figure*}

We have shown in the previous section that a large sample of
gravitational lens systems is likely to contain a significant
fraction of double lens systems.  It is instructive to look at the
properties of the double lens pairs in the sample.\\
Fig.\,\ref{masses} shows the contribution to the total lensing cross section
as a function of the ratio of the two individual lens masses in panel (a) and as a function of the
ratio of the two lens redshifts in panel (b). It can be seen that most of the double
lens pairs in this sample have lenses lying at similar redshifts and
with similar masses. 
This is not surprising, as the cross section for lensing of sources
at a given redshift peaks at an optimal value of the redshift and mass
of the lens galaxy. Positive spatial
correlation will increase the number of double lens systems that
are likely to lie at similar redshifts further.
Note, however, that for uncorrelated pairs the efficency of lensing as a
function of redshift is far
less peaked for the second lens than that for a single lens (see, for example, Fig.\,5 in
Fukugita et al. 1992). In fact, the likelihood that the ratio of redshifts
of uncorrelated double lenses is 2 or larger is $\sim50$ per cent.

\subsection{Statistical image geometries}

\begin{figure*}
\epsfig{file=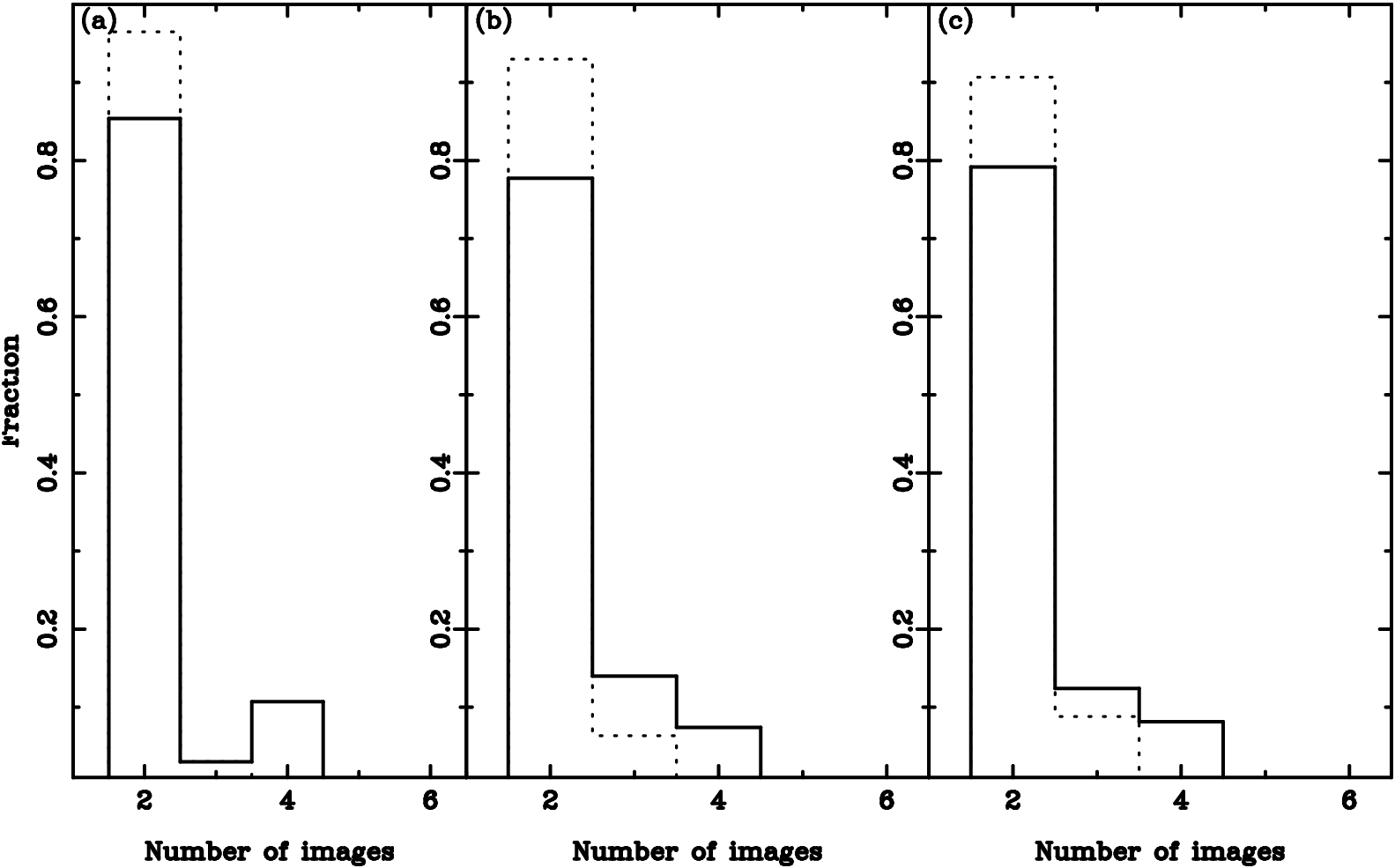,width=8.0cm,angle=-90}
\caption{Number of images for strong and weakly coupled double lens systems. The solid line
marks the histogram for strongly coupled double lenses in which the caustics merge. The
dotted line marks the histogram for weakly coupled double lens systems in which a second lens introduces
significant external shear but for which the caustics do not
merge. The statistical uncertainty on all results shown in this figure is
about 5 per cent. The source redshift is $z_{\mathrm{s}}=2$.}
\label{imhist} 
\end{figure*}

Image geometries and magnification bias are strongly dependent on the
mass profile of the lens, and so a two lens system is expected to lead
to unusual and complex image geometries. In Section 2, we showed some
individual image geometries for extended sources.  To determine the
statistical properties of the images, we generated images for small
sources in the double lenses in our sample. The histogram of the cross
section for the number of images is shown in Fig.\,\ref{imhist}.
We show the results for two different sets of observational selection criteria. In panel (a) the statistics include all images, in panel (b) only images with minimum separation of $0.05\arcsec$ and a maximum magnification ratio of 100 are included and in panel (c) only images with minimum separation of $0.05\arcsec$ and maximum magnification ratio of 20 are included. The
histograms show clearly the increased cross section towards three
($\sim15\%$) and four ($\sim10\%$) images. 
Individual spherical PIMD lenses can only produce two magnified and one demagnified image.  Strongly coupled double lens systems can produce three or more magnified images. Strongly
and weakly coupled double lenses both lead to quadruple systems and
to a small fraction of five- and six-image systems. However, since the
fraction of double lens systems will be small, and the vast majority of
all lens galaxies are expected to have some effective elliptical profile,
which increases the cross section for the formation of four or more images in a
similar way, the effect of double lenses on the overall image
statistics in a large sample of lenses will be small. Note that the image 
configuration shown in Fig.\,\ref{mags3}(b) is characteristic of double lens
systems and hard to reproduce in most single lens models. Also, three
strongly magnified images, without an additional counter-image, is a
clear sign of a possible double lens system. 
Magnification bias in any flux limited lens sample will increase the
relative number of lens systems with high image magnifications and
hence high image multiplicities. This effect will increase the
fraction of systems with 3 or more images by a few per cent.

\section{Extending double lens models}

\subsection{Double lens plus external shear}
\begin{figure*}
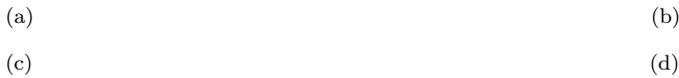

\begin{minipage}{170mm}
(a) \hskip 81mm (b)
\begin{center}
\vskip -5mm
\hskip 9mm
\end{center}
(c) \hskip 81mm (d)
\begin{center}
\vskip -8mm
\hskip 9mm
\end{center}
\caption{Magnification maps and images for double lenses similar to
those in Figs.\,\ref{mags1}-\ref{mags4} with an additional component of external
shear. The lenses have the same parameters as those in Fig.\,\ref{mags3} and
are at $z_{\mathrm{A}}=0.6$ and $z_{\mathrm{B}}=0.3$ respectively with an angular separation
of $3.0''$. Panels (a) and (c) show the source plane, panels (b) and
(d) the image plane. The shear is due to an external point lens of
mass $7\times10^{12}\mathrm{M}_{\odot}$ at a distance of 14\arcsec at a
redshift of 0.2. The mass lies in the lower left along the line
connecting the lenses in the lower panels and in the bottom right along
a line perpendicular to that connecting the two lenses in the upper panels.}
\label{shear}
\end{minipage}
\end{figure*}

Our above analysis does not assume any correlation between
the lens positions, however, the spatial positions of galaxies on the
sky are in reality correlated, and most galaxies occur
in small groups or clusters. To show qualitatively how
other galaxies in the environment affect double lensing, we model the
potential perturbation due
to a nearby group or cluster as an
external shear that acts on the double lens system. In
Fig.\,\ref{shear} we show the magnification maps for the double lens
system shown in Fig.\,\ref{mags3}, now modified by an external shear,
which is assumed to be perpendicular to the alignment of the lenses in
Fig.\,\ref{shear}(a) and parallel to the alignment of the lenses
in Fig.\,\ref{shear}(b). 
We model
the source of the external shear as a point mass of $7\times 10^{12}\,
\mathrm{M}_{\odot}$ at a distance of 14 arcsec. In Fig.\,\ref{shear}(a) the
external shear ``bends'' the caustic structures in its
direction, but the effect on the caustic area is small. It is only when
the external perturber lies close to and is nearly perfectly
aligned with the lens pair, as in Fig.\,\ref{shear}(c), that the size of the caustic is affected
significantly. Such a situation is likely to be rare, however, and so we expect
that, on average, the effect on lensing statistics due to external
shear from perturbers is likely to be small. However image positions and magnifications can be
affected greatly by even 
moderate external shear, and so external perturbers have to be taken
into account in accurate lens modelling of individual double lens systems. A more
detailed treatment of the effect of external perturbers like groups of
galaxies on the properties of galaxy lenses can be found in Keeton, Kochanek \& Seljak (1997) and M\"oller, Natarajan \& Kneib (in preparation).

\subsection{Spiral galaxies}
\begin{figure*}
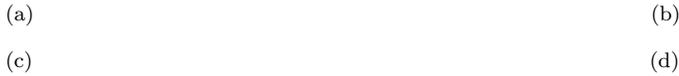

\begin{minipage}{170mm}
(a) \hskip 81mm (b)
\begin{center}
\vskip -5mm
\hskip 9mm
\end{center}
(c) \hskip 81mm (d)
\begin{center}
\vskip -8mm
\hskip 9mm
\end{center}
\caption{Magnification maps and images for spiral double lenses. Panels (a) and (c) show the source plane, panels (b) and (d)
the image plane. The lenses are both modelled as the sum of a PIMD and
an exponential disc. The lens separation and PIMD halo properties are as in Fig.\ref{mags3}. The discs have a surface
mass density of $\Sigma_0=10^{10}\mathrm{M}_{\odot}\mathrm{kpc}^{-2}$ and a scale length of
$r_{\mathrm{s}}=3\,\mathrm{kpc}$. The lens in the top right of each panel is inclined at 75
deg towards the line of sight and is at $z_1=0.6$, whereas the lens in
the bottom left of each panel is inclined at 65 deg to the line of sight
at $z_2=0.3$. In the upper panels, the discs are aligned towards each
other, in the lower panels they are aligned at right angles to the line
connecting the two lenses.}   
\label{discs}
\end{minipage} 
\end{figure*}

In the above analysis we kept the number of parameters as
small as possible and used the PIMD as a simple but
reasonable lens model. Including
spiral discs into the lens model significantly changes the lensing
behaviour of individual systems, as shown by Maller, Flores \& Primack
(1997) and MB98. The statistical lensing properties of
spiral lenses has been investigated by Bartelmann \& Loeb (1998), Keeton \& Kochanek (1998) and Blain, M\"oller \& Maller
(1999). These studies showed that there is a
significant effect on the properties of lensing by a single galaxy. We show
qualitatively how the presence of spiral discs will affect double lensing in
Fig.\,\ref{discs}. 
Both lenses are spiral galaxies similar to the
Milky Way, which consists of an PIMD halo and an inclined thin exponential disc, of central surface mass density $\Sigma_0$ and scale length $r_{\mathrm{s}}$,
containing about 10 per cent of the halo mass. The effect of the discs
on the lens properties depends strongly
on the relative alignment of the discs. If they are perpendicular to
one another, then the shear along the caustics is greatly reduced and the
area enclosed by the caustics shrinks drastically. If both discs are aligned with each other and the angle between their major axes is less than about 60 deg, then the caustic lines can enclose a very
large high--magnification area as seen in Fig.\,\ref{discs}(a). It is difficult to assess the effect of
spirals on double lens systems in a statistical sense due to the large
parameter space. The effect of
discs on double lens statistics is expected to be small if the alignment of the discs is uncorrelated,
as the asymmetry that is
introduced in the potential by the disc will \emph{on average} counteract the
ellipticity introduced due to a second lens. Individual double spiral
galaxy lenses will in general have different properties than double
spherical or elliptical lenses. Spiral galaxies are at least twice as abundant as elliptical galaxies
and therefore, since the inclination effect due to the discs enhances the high magnification cross section, double lens systems containing one or two spiral galaxies
are likely. Observations of such lens systems could provide strong
constraints on the mass profile of one or both lenses.

\section{Observations of double lenses}

The observed properties of double lenses are likely to depend more strongly on
the wavelength of observation than those of single
lenses. This is due to the fact that light from
the images and the more distant lens (or its images if it is
multiply lensed by the nearer lens) is likely to be
superimposed on the light distribution of the near lens galaxy. In
addition, dust in the interstellar medium of the lens galaxies could
lead to significant extinction of one or more of the images.
Double lens
systems will only be easily observable if both lenses are
relatively faint at the wavelength of
observation. This
would be the case, for example, in the radio or sub-mm
waveband if the two lenses are elliptical galaxies and the background
source is a small young star forming galaxy.
It would not be the case for optical observations of two massive
spiral galaxies lensing a distant galaxy. In this case images are
likely to be too faint
relative to the lenses to be observable at most wavelengths. As quasar
radio surveys are
not affected by extinction, a complete lens sample from a radio
survey, like the CLASS survey, would therefore be especially suited to
observe double lens systems. The {\it Planck} mission will discover many thousands distant
sources serendipitously at sub-mm wavelengths of which a fraction of
order 10 per cent could 
be lensed (Blain 1998a). We predict that about 5 per cent of these
lens systems will be double lenses. The Atacama Large Millimetre Array (ALMA) which will observe
at sub-mm wavelengths would be well suited to detect the images
of distant sources (Blain 1998b;2000). The subarcsecond resolution of ALMA would resolve the individual images and allow their direct study, which would greatly improve the accuracy of lens mass models.

\section{Conclusions}

In this paper we have investigated gravitational lensing of high-redshift
background sources by more than one galaxy along the line of
sight. Using a Press--Schechter halo distribution and a ray-tracing code we
have estimated the number of double lens systems that are to be
expected in a large lens sample. We have discussed the properties of such double
lens systems and investigated more complicated double lens models qualitatively. 
In summary, our main results are:
\begin{enumerate}
\item In a cosmology with $\Omega_{\Lambda}=0.7$ and
$\Omega_{\mathrm{M}}=0.3$, about 2-5 per cent of all
multiply imaged sources at $z\approx2$ or higher are expected to be
lensed by more than one lens along the line of sight.
\item The second lens induces a strong asymmetry in the effective
lensing potential. This leads to a significant change in the caustic
structure. The cross section for high magnification of point sources
increases significantly due to this effect. 
\item Double lenses lead to a significant fraction of lens systems
with three
($\sim15\%$) and four ($\sim10\%$) images, and can lead to five- and six-image configurations. 
\item The two lenses in a double lens system are likely to
be of similar mass and redshift. 
\item Additional external shear acting on a lens pair can modify the
caustic structure and the image geometries in individual systems significantly, and needs to be included
in lens modelling just as for single lens systems.
\item Double spiral galaxy lenses can have a large high-magnification
cross section, if their postion angles are aligned with each other, and
the inclination of both of the discs towards the line of sight is higher than
about 65 deg.
\item Future lens surveys, especially in the submillimetre wavebands will contain a significant number of double lens systems.
\end{enumerate}
\section*{Acknowledgements}

OM acknowledges the Lensnet TMR network and AWB the Raymond and Beverly Sackler Foundation for support.
We thank Priya Natarajan, Daniel Mortlock and the referee, Jean-Paul Kneib for useful comments on the manuscript.


\begin{thebibliography}{}

\bibitem[\protect\citename{Bartelmann \& Loeb, 1997}]{CLOVER}
Bartlemann M., Loeb A., 1998, ApJ, 503, 48
\bibitem[\protect\citename{bl}%
]{B96}
Blain A.\,W., 1996, MNRAS, 283, 1340
\bibitem[\protect\citename{bl}%
]{B96}
Blain A.\,W., 1998a, MNRAS, 295, 92
\bibitem[\protect\citename{bl}%
]{B96}
Blain A.\,W., 1998b, MNRAS, 297, 511
\bibitem[\protect\citename{bl}%
]{B96}
Blain A.\,W., 2000, in Wootten A., ed., Science with the Atacama Large
Millimeter Array, PASP conf series, in press (astro-ph/9911449)
\bibitem[\protect\citename{bl}%
]{B96}
Blain A.\,W., M\"oller, O., Maller, A.\,H. 1999, MNRAS, 303, 423 
\bibitem[\protect\citename{Falco et al., 1999}]{CASTLES}
Falco E.\,E., et al.,2000, in Brainerd T. \& Kochanek C.\,S., eds.,
Gravitational Lensing: Recent Progress and Future Goals, ASP, San Francisco, in press (astro-ph/9910025) 
\bibitem[\protect\citename{Helbig et al., 1999}]{CLOVER}
Helbig P., Marlow D.\,R. Quast R., Wilkinson P.\,N., Browne
I.\,W.\,A., Koopmans L.\,V.\,E. 1999, A\&A Supp, 136, 297
\bibitem[\protect\citename{M\"oller}]{MOELL}
Jain B., Seljak U., White S., 2000, ApJ, 530, 547   
\bibitem[\protect\citename{Keeton \& Kochanek}]{JAUNSEN}
Keeton C.\,R., Kochanek C.\,S., 1997, ApJ, 487, 42
\bibitem[\protect\citename{Keeton \& Kochanek}]{JAUNSEN}
Keeton C.\,R., Kochanek C.\,S., 1998, ApJ, 495, 157
\bibitem[\protect\citename{Keeton \& Kochanek}]{Keeton}
Keeton C.\,R., Kochanek C.\,S., Seljak U., 1997, ApJ, 482, 604
\bibitem[\protect\citename{Kneib et al., 1997}]{CLOVER}
Kneib J.-P., Alloin D., Mellier Y., Guilloteau S., Barvainis R.,
Antonucci R., 1998, A\&A, 329, 827
\bibitem[\protect\citename{Kneib et al., 1997}]{CLOVER}
Kneib J.-P., Cohen J.\,G., Hjorth J., 2000, ApJL, submitted (astro-ph/0006106)
\bibitem[\protect\citename{Kneib et al., 1997}]{CLOVER}
Kneib J.-P., Ellis R.\,S., Smail I., Couch W.\,J., Sharples R.\,M., 1996, ApJ, 471, 643
\bibitem[\protect\citename{Kochanek, 1988}]{CLOVER}
Kochanek C.\,S., Apostolakis J., 1988, MNRAS, 235, 1073
\bibitem[\protect\citename{Koopmans, de Bruyn \& Jackson}]{KOOP}
Koopmans L.\,V.\,E., et al., 1999, MNRAS, 303, 727
\bibitem[\protect\citename{Koopmans, de Bruyn \& Jackson}]{KOOP}
Koopmans L.\,V.\,E., de Bruyn A.\,G., Jackson N., 1998, MNRAS, 295, 534
\bibitem[\protect\citename{Koopmans, de Bruyn \& Jackson}]{KOOP}
Koopmans L.\,V.\,E., Fassnacht C.\,D., 1999, ApJ, 527, 513
\bibitem[\protect\citename{Koopmans, de Bruyn \& Jackson}]{KOOP}
Kundi\'c T., et al., 1997, AJ, 114, 2276
\bibitem[\protect\citename{Lewis}]{KOOP}
Lewis G.\,F., Miralda-Escude J., Richardson D. C., Wambsganss J., 1993, MNRAS, 261, 647
\bibitem[\protect\citename{Maller, Flores \& Primack, 1997}]{MAL}
Maller A.\,H., Flores R.\,A., Primack J.\,R., 1997, ApJ, 486, 681
\bibitem[\protect\citename{M\"oller}]{MOELL}
M\"oller O., 1997, MSci Dissertation, Cambridge University
\bibitem[\protect\citename{M\"oller}]{MOELL}
M\"oller O., Blain A.\,W., 1998, MNRAS 299, 845 
\bibitem[\protect\citename{M\"oller}]{MOELL}
M\"oller O., Natarajan P., 2000, in Brainerd T., Kochanek C.\,S., eds.,
Gravitational Lensing: Recent Progress and Future Goals, ASP, San Francisco , in press
(astro-ph/9909303).  
\bibitem[\protect\citename{Myers et al. 1995}]{MYERS}
Myers S.\,T. et al., 1995, ApJ, 447, L5
\bibitem[\protect\citename{Natarajan \& Kneib 1997}]{NAT97}
Natarajan P., Kneib J.-P., 1997, MNRAS, 287, 833
\bibitem[\protect\citename{Peebles, 1993}]{CLOVER}
Peebles P.\,J.\,E., 1993, Principles of Physical Cosmology, Princeton University Press, Princeton, p.213
\bibitem[\protect\citename{Pei, 1995}]{PEI95}
Pei Y.\,C., 1995, ApJ, 440, 485
\bibitem[\protect\citename{Press et al. 1988}]{PRESS}
Press W.\,H. Flannery B.\,P., Teukolsky S.\,A., Vetterling W.\,T., 1988, Numerical Methods in C, Cambridge University Press
\bibitem[\protect\citename{Press \& Schechter, 1974}]{CLOVER}
Press W.\,H., Schechter P., 1974, ApJ, 187, 425
\bibitem[\protect\citename{Schneider et al., 1992}]{SCHNEID}
Seitz S., Schneider P., 1992, A\&A, 265, 1
\bibitem[\protect\citename{Schneider et al., 1992}]{SCHNEID}
Schneider P., Ehlers J., Falco E.\,E., 1992, Gravitational Lenses, Springer, New York
\bibitem[\protect\citename{Schneider et al., 1992}]{SCHNEID}
Soucail G., Kneib J.-P., Jansen A.\,O., Hjorth J., Hattori M., Yamada T., 2000, A\&A, submitted (asto-ph/006382)
\bibitem[\protect\citename{Wambsganss et al., 1997}]{CLOVER}
Wambsganss J., Cen R., Ostriker J.\,P., 1998, ApJ, 494, 29
\bibitem[\protect\citename{Wambsganss et al., 1997}]{CLOVER}
Wambsganss J., Witt H.\,J., Schneider P., 1992, A\&A, 258, 591
\end{thebibliography}
\end{document}